\documentclass[a4paper,11pt]{article}
\usepackage{jcappub} 
\usepackage{lineno}
\usepackage{xcolor}
\usepackage{aas_macro}

\title{\boldmath 
Massive $\nu$s through the CNN lens: \\
interpreting the field-level neutrino mass information in weak lensing

}

\author[a,b]{Malika Golshan}
\author[c,d]{and Adrian E.~Bayer}

\affiliation[a]{
Department of Physics, University of California, Berkeley, CA 94720, USA}%
\affiliation[b]{Lawrence Berkeley National Laboratory,  1 Cyclotron Road, Berkeley, CA 94720, USA}%
\affiliation[c]{
Department of Astrophysical Sciences, Princeton University, Peyton Hall, Princeton NJ 08544, USA}%
\affiliation[d]{
Center for Computational Astrophysics, Flatiron Institute, 162 5th Avenue, New York NY 10010, USA}%

\emailAdd{malikagolshan@berkeley.edu}
\emailAdd{abayer@princeton.edu}

\abstract{
Modern cosmological surveys probe the Universe deep into the nonlinear regime, where massive neutrinos suppress cosmic structure. Traditional cosmological analyses, which use the 2-point correlation function to extract information, are no longer optimal in the nonlinear regime, and there is thus much interest in extracting beyond-2-point information to improve constraints on neutrino mass. 
Quantifying and interpreting the beyond-2-point information is thus a pressing task.
We study the field-level information in weak lensing convergence maps using convolution neural networks.
We find that the network performance increases as higher source redshifts and smaller scales are considered -- investigating up to a source redshift of 2.5 and $\ell_{\rm max}\simeq10^4$ -- verifying that massive neutrinos leave a distinct effect on weak lensing. However, the performance of the network significantly drops after scaling out the 2-point information from the maps, implying that most of the field-level information can be found in the 2-point correlation function alone.
We quantify these findings in terms of the likelihood ratio and
also use Integrated Gradient saliency maps to interpret which parts of the map the network is learning the most from. We find that, in the absence of noise, the network extracts a similar amount of information from the most overdense and underdense regions. However, upon adding noise, the information in underdense regions is distorted as noise disproportionately washes out void-like structures.
}

\begin{document}
\maketitle
\flushbottom

\section{Introduction}
\label{sec:intro}

Neutrinos have historically been among the most challenging elementary particles to detect and study, primarily because they interact only through the weak force and  gravity. While their gravitational interaction is very weak due to their small mass, cosmological experiments are able to infer the neutrino mass from the distribution of radiation and matter in the Universe. Analysis of cosmic microwave background (CMB) data from Planck combined with large-scale structure (LSS) data from BOSS have presented upper limits on the neutrino mass of  $M_\nu < 0.12$ (95\% CL) \citep{collaboration2018planck}. This is an order of magnitude lower than the upper limit on the effective electron neutrino mass measured in particle physics experiments of $\beta$ decay which give $m_{\nu_{e}}^{\rm eff} \lesssim 1.1 {\rm eV}$ \citep{Aker_2019}. 
Given the minimum sum of the neutrino masses allowed by neutrino oscillations experiments \citep{SuperK, SNO, KamLAND, K2K, DayaBay} is $M_\nu \gtrsim 0.06 {\rm eV}$ for the normal hierarchy, or $M_\nu \gtrsim 0.1 {\rm eV}$ for the inverted hierarchy, there is much promise for the next generation of cosmological surveys -- such as Vera Rubin Observatory LSST~\cite{LSST}, Euclid~\cite{Euclid}, SPHEREx~\cite{SPHEREx:2018xfm}, Roman Space Telescope~\cite{RST}, DESI~\cite{DESI}, PFS~\cite{PFS}, Simons Observatory~\cite{SimonsObservatory:2018koc}, CMB-S4~\cite{CMB-S4:2016ple} -- to measure the neutrino mass at the $5\sigma$ level.

One particularly promising method to measure neutrino mass is by studying weak gravitational lensing. This involves analyzing the distortions of light emitted from distant galaxies caused by the gravitational influence of LSS between the galaxy and the telescope. Traditionally, information is extracted from weak-lensing convergence maps using the power spectrum, or 2-point correlation function, which is optimal in the linear regime. However, upcoming surveys will probe progressively smaller, nonlinear, scales, in which regime the 2-point statistics are no longer guaranteed to be optimal.
Various simulations have been developed to understand the effects of neutrinos on nonlinear scales
\cite{Liu2018MassiveNuS:Simulations, Bird2018, Villaescusa_Navarro_2018, Arka_2018, Bayer_2021_fastpm, DeRose:2023dmk, Sullivan:2023ntz}, and have suggested that there is much information to be found on these nonlinear scales using higher-order statistics such as voids and wavelets. 
For example, when considering the 3-dimensional matter field it has been shown that higher-order statistics can improve neutrino mass constraints by a factor of $\sim100$ compared to the power spectrum alone \cite{Hahn_2020, hahn2020constraining, Uhlemann_2020, Massara_2020, Valogiannis:2021chp, bayer2021detecting, Kreisch_2021, Bayer:2024xfb}, while for weak lensing more modest numbers are reported \cite{Kreisch2019,liu&madhavacheril2019, Li2019, Coulton2019, Coulton2020, Marques2019, Boyle:2020bqn, ajani2020, Ajani:2021pgp, Cheng_2021}. 
Understanding and interpreting the source of this information is thus a pressing task.

In this work, we extend the analysis of \cite{Bayer_2022_fake}, which showed that while there is much information regarding neutrino mass in the 3-dimensional matter field, almost all of the information up to $k\sim1\,h/{\rm Mpc}$ is contained in the power spectrum alone for observable fields, such as the weak lensing convergence field and galaxy field. 
This conclusion was reached by considering the 3-dimensional field-level cross correlation between two simulations with matched linear physics and different neutrino masses at a single redshift, essentially assuming the distance between lensing source and observer to be small. As will be discussed, we perform an analysis using convolutional neural networks (CNNs) with ray-traced N-body simulations and reach a similar conclusion. 

While higher-order statistics provide an idea of the information content in a cosmic field, (i) they are not necessarily optimal -- perhaps there is much more information to be found -- and (ii) it is possible that much of the information in the higher-order statistics is also available in the 2-point correlation function. 
In this work we address both of these points by (i) using a CNN to extract all the information at the field-level and (ii) studying how much information there is beyond the 2-point by scaling out the 2-point information from the maps before training the CNN. 

CNNs have been employed in many studies to estimate cosmological parameters in weak lensing and other areas of cosmology, generally giving tighter constraints than other higher-order statistics \citep{Gupta_2018, Ribli_2019, Fluri_2018, Fluri_2019, villaescusanavarro2020neural,Jeffrey_2020,Matilla_2020, Makinen_2021, Lu_2022, Lu_2023, SimBIG:2023ywd, Sharma:2024pth}, but have yet to be applied in the context of neutrino mass. 
Here, we train a classifier to distinguish between simulations with neutrino masses of $M_\nu = 0$eV and $0.1$eV (the minimum value allowed for the inverted hierarchy). 
We perform classification instead of full parameter inference due to the lack of currently available weak lensing simulations for neutrino mass parameter inference.
We study the performance of the network as a function of maximum scale $\ell_{\rm max}$, source redshift $z$, and noise level $\sigma_\kappa$. 

Machine learning models are often used as black boxes, but with the wide use of neural networks in astrophysics and cosmology there has been a recent drive to interpret and explain what the networks are learning \cite{Lucie-Smith:2018smo, Lucie-Smith:2019hdl, Lucie-Smith:2020ris, Lucie-Smith:2023kue, ZorrillaMatilla:2020doz, Bhambra_2022, you2023sumofpartsmodelsfaithfulattributions}. 
We study Integrated Gradient saliency maps \cite{sundararajan2017axiomatic} to determine which parts of weak lensing convergence maps are most informative about neutrino mass. 
This is of particular interest as there is much discussion about underdense regions of cosmic structure, known as voids, containing a large amount of information regarding neutrino mass in the context of 3-dimensional LSS \cite{Kreisch_2020, bayer2021detecting, Kreisch_2021, Bayer_2022_fake, Thiele:2023oqf, Bayer:2024xfb}.

The paper is organized as follows. Section \ref{sec:method} describes the simulation data and neural network, Section \ref{sec:results} reports our results, and Section \ref{sec:conclusions} provides conclusions and discussion.

\section{Method}
\label{sec:method}

In this section, we describe the data, neural network architecture, and analysis methods.

\subsection{Data}

\begin{figure*}
  \centering
  \includegraphics[width=\textwidth]{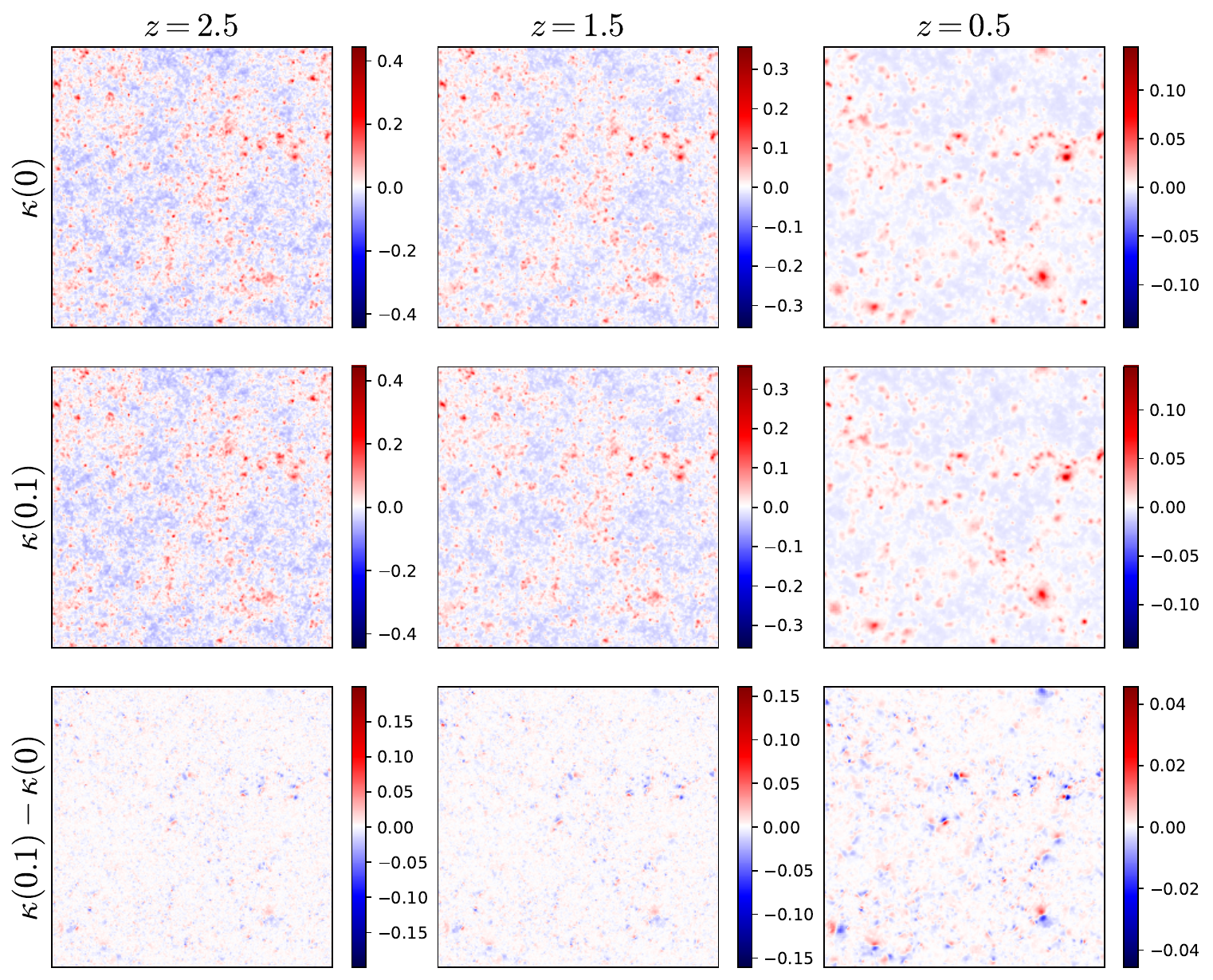}
  \caption{A particular realization of convergence maps $\kappa$ from \texttt{MassiveNuS}. The first row shows the maps for a 0eV cosmology, the second row for a seed-matched 0.1eV cosmology, and the final row shows the difference between the two. Different columns show different source redshifts.}
  \label{fig:maps}
\end{figure*}

We use the \texttt{MassiveNuS} simulations \cite{Liu2018MassiveNuS:Simulations}, which are based on a modified version of the N-body code \texttt{Gadget-2}~\citep{Gadget}. We use a set of $10{,}000$ pairs of 
simulations with $M_\nu=0\rm{eV}$ and $0.1\rm{eV}$. The other cosmological parameters are fixed to $A_s = 2.1 \times 10^{-9}$, $\Omega_m=0.3$, $\Omega_b = 0.05$, and $h = 0.7$.
Each simulation contains $1024^3$~particles in a  ($512\,{\rm Mpc}/h$)$^3$ box, thus the Nyquist wavenunber is $k_{\rm Nyq}=2\pi\,{h/{\rm Mpc}}$. Weak lensing convergence maps $\kappa$ are produced by ray-tracing using the flat-sky approximation over a $(3.5 \,{\rm deg})^2$ solid angle. We consider three source redshifts of $z=2.5,1.5,0.5$. Figure \ref{fig:maps} shows a particular realization of convergence maps; the first row shows a map for the 0eV cosmology, the second row for a seed-matched 0.1eV cosmology, and the final row shows the difference between the two. It can be seen that there is stronger, more nonlinear, structure for larger source redshifts -- in particular there are sharper overdense regions and larger underdense regions, leading to sharper differences caused by the difference in neutrino mass, particularly in overdense regions.

\begin{figure*}
  \centering
  \includegraphics[width=0.7\textwidth]{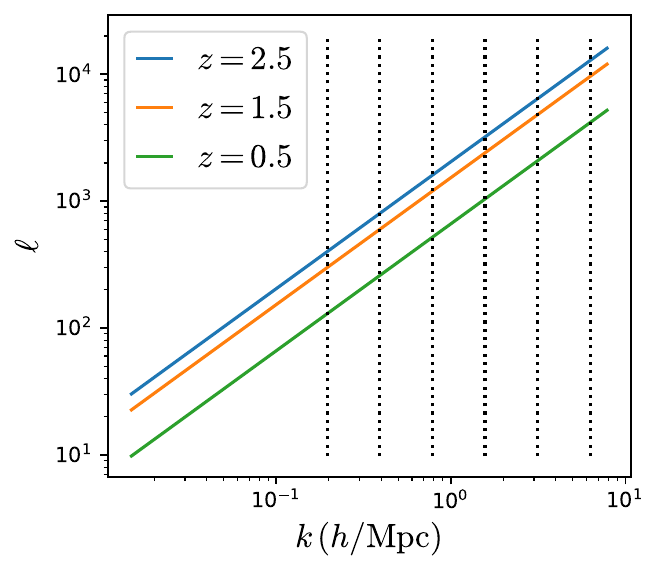}
  \caption{The relation between 2-dimensional wavevector $k$ and multipole $\ell$ for the source redshifts considered in this work. The dotted lines represent the various $k_{\rm max}$ cuts we perform in the analysis.}
  \label{fig:k2l}
\end{figure*}

Before feeding the data to the CNN we apply augmentations for various purposes:
\begin{enumerate}
    \item \textbf{Cutting in Fourier Space}: The images were transformed to Fourier space and cut at various scales in factors of 2 of the Nyquist, $k_{\rm max}=2\pi,\pi,\pi/2,\pi/4,\pi/8,\pi/16\,({h/{\rm Mpc})}$, and then transformed back into configuration space. As the maps are produced using the flat-sky approximation, we use a 2-dimensional wavenumber, denoted $k$, to represent the scale. To connect with curved-sky intuition, we show the relation between $k$ and multipole $\ell$ in Figure \ref{fig:k2l} for the various source redshifts considered.
    \item \textbf{Rescaling to remove the 2-point information}: As well as considering the information in the convergence maps, we additionally analyse the maps after rescaling the 0eV maps to fake the effect of neutrinos at the level of the power spectrum. We create the maps with fake $\nu$s at the 2-point level by multiplying all of the 0eV convergence maps by the square-root of the ratio of the 0.1eV to 0eV power spectra as follows: $\kappa(0) \rightarrow \sqrt{\frac{P(0.1)}{P(0)}}\kappa(0)$, where $P(M_\nu)$ is the 2-dimensional power spectrum averaged over all simulations with neutrino mass $M_\nu$. This effectively matches the 2-point information in the 0eV and 0.1eV cosmologies and enables study of the beyond 2-point information.
    \item \textbf{Addition of Gaussian Noise}: Gaussian noise with zero mean was added to mimic observational noise. We considered two standard deviations of $\sigma_\kappa=0.01$ and 0.1 (in the same units as $\kappa$), which is around the order of magnitude of noise expected for an LSST-like survey with a galaxy number density of $n_{\rm gal}\simeq 5\,{\rm arcmin}^{-2}$ (for one tomographic bin) and a shape noise of $\sigma_\lambda\simeq0.3$ \cite{liu&madhavacheril2019}. We also consider the noiseless maps.
    \item \textbf{Normalization}: The maps were normalized to have zero mean and unit standard deviation.
    \item \textbf{Data Augmentation}: The images were randomly flipped and rotated to increase the variability and robustness of the training set.
\end{enumerate}

\subsection{Neural Network}

We use a ResNet-18 network \cite{he2015deepresiduallearningimage} to perform the analysis. ResNet architectures have commonly been used to achieve start-of-the-art performance on various computer-vision deep-learning tasks over the last few years.
ResNets are a group of CNN models made to skip one or more layers to allow for the network to learn residual functions with reference to the layer input rather than to unreferenced functions. This is achieved via skip, or residual, connections, and enables training a network with a large depth. The architecture of the model starts with an initial convolutional layer followed by a max-pooling layer, which reduces the spatial dimensions of the input image. This is followed by a series of residual blocks grouped into four stages, each containing two residual blocks with increasing filter sizes. Each residual block consists of two convolutional layers, batch normalization and ReLU activation, with shortcut connections that add the input of the block to its output.  The final stages of the network include a global average pooling layer and a fully connected layer that outputs the class probabilities.
We choose a ResNet-18 as \cite{Sharma:2024pth} reported this to be sufficient for weak lensing cosmology applications, finding that using a deeper ResNet, or an architecture with a larger receptive field, did not improve the performance.

We modified the ResNet-18 for our applications to take only 1 channel of input, instead of the default 3. We also replaced the final layer of the network by 2 fully connected layers---each followed by dropout and a LeakyReLU activation function---and then a third fully connected layer to output a scalar via a sigmoid activation function. We then apply a binary cross entropy loss function to train a classifier between the 0eV and 0.1eV neutrino mass simulations. For optimization we used the ADAM optimizer with an exponential learning rate scheduler, using early stopping to avoid overfitting. We report results for a batch size of 32, but also found that using a batch size of 64 or 128 made no improvement to the model. We use the pre-trained ResNet-18 weights as the starting point for training, but then refit all the weights during training. 
The data set was divided into training, validation, and testing sets according to a 70:15:15 split.

We trained the neural network for all the different source redshifts, scale cuts, noise levels, and power spectrum augmentations detailed in the previous subsection. 
We used \texttt{optuna} \cite{optuna} to optimize the hyperparameters of the network and training, in particular optimizing the dropout rate, weight decay, learning rate, and learning rate scheduler decay rate. We performed 200 trials for every training and took the model(s) with the highest validation accuracy.

Subsequently, we applied an interpretability metric, the Integrated Gradient saliency map \cite{sundararajan2017axiomatic}, to identify which pixels in the map the CNN is paying the closest attention to. The Integrated Gradient quantifies how the model output (the neutrino mass) changes with respect to the model inputs (the pixel values) by propagating the gradient through the network parameters. This allowed us to determine which parts of the maps contains the most information regarding neutrino mass. We implemented the saliency map using the \texttt{Captum} library \cite{captum}.

\section{Results}
\label{sec:results}

\begin{figure*}[t]
  \centering
  \includegraphics[width=\linewidth]{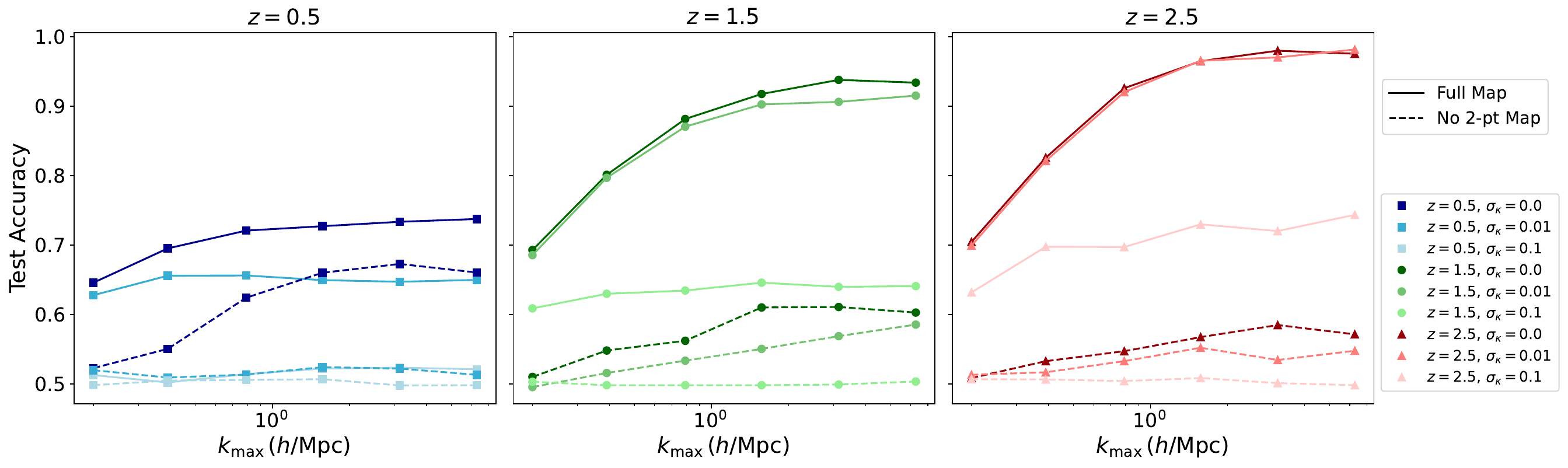}
  \includegraphics[width=\linewidth]{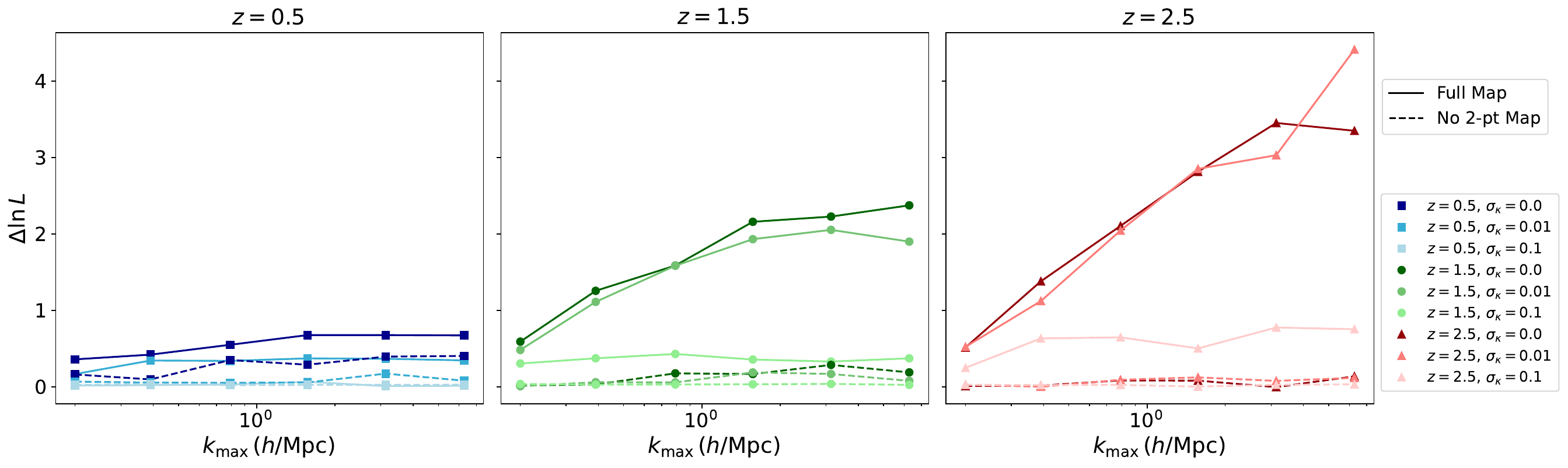}
  \caption{\textit{Top:} classification accuracy evaluated on the test set as a function of $k_\text{max}$. \textit{Bottom:} difference in log likelihood between a cosmology with $M_\nu=0.1$ and $0\,{\rm eV}$ for observations with $M_\nu=0.1\,{\rm eV}$ (taken from the test set) as a function of $k_\text{max}$.
  Each panel and base color corresponds to a different source redshift $z$. The difference in color shades corresponds to the noise $\sigma_\kappa$, with lighter shades corresponding to higher noise. Solid lines represent the accuracy for a model trained on the full convergence maps, while dashed lines are for maps that have had the 2-point information scaled out. It can be seen for full maps that the accuracy and difference in likelihood are typically higher at smaller scales, higher redshifts, and lower noise. However, when the 2-point information is scaled out the accuracy and difference in likelihood are consistently significantly worse -- almost $50\%$ and 0 respectively -- indicating that much of the field-level information is contained in the 2-point alone.} 
  \label{fig:valid_acc}
\end{figure*}

\begin{figure*}[t]
  \centering
  \includegraphics[width=\textwidth]{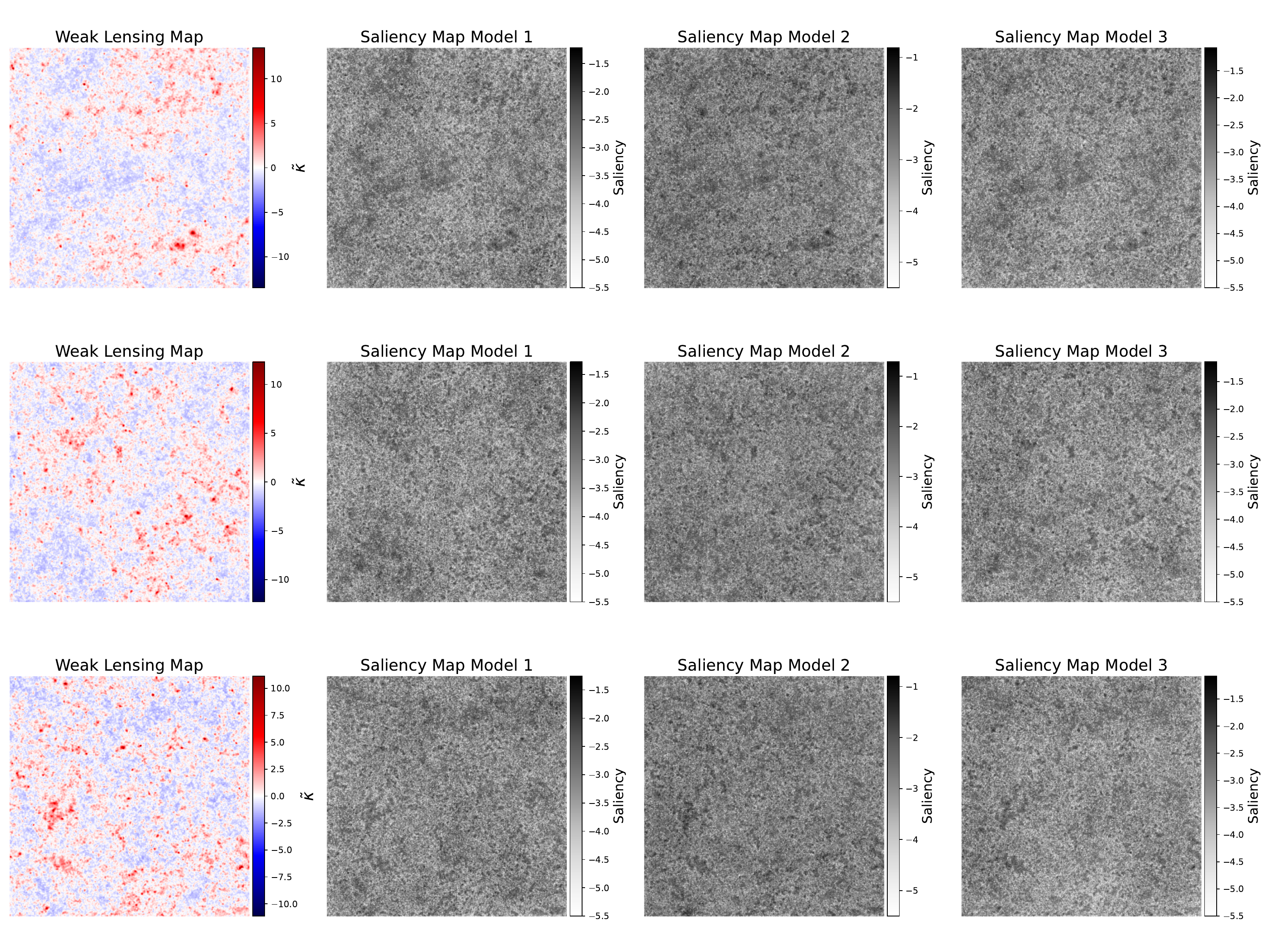}
  \caption{Visualization of the logarithm of the absoulte value of the Integrated Gradient saliency maps for the case of $z= 2.5$, $k_{\text{max}}= {\pi}/{2}$, and $\sigma_\kappa = 0.01$ (without scaling out the 2-point information). Each row displays a different (mean-subtracted and variance-scaled) weak lensing convergence map in the left column, followed by three saliency maps generated by the best three models from the \texttt{optuna} trials. Higher numbers in the saliency map implies that the network is paying more attention to that region of the map. The saliency maps show a wide variability in terms of regions of the map they focus on, in some cases focusing on underdense regions, and in other cases focusing on overdense regions.}
  \label{fig:saliency}
\end{figure*}

\begin{figure*}[t]
  \centering
  \includegraphics[width=\textwidth]{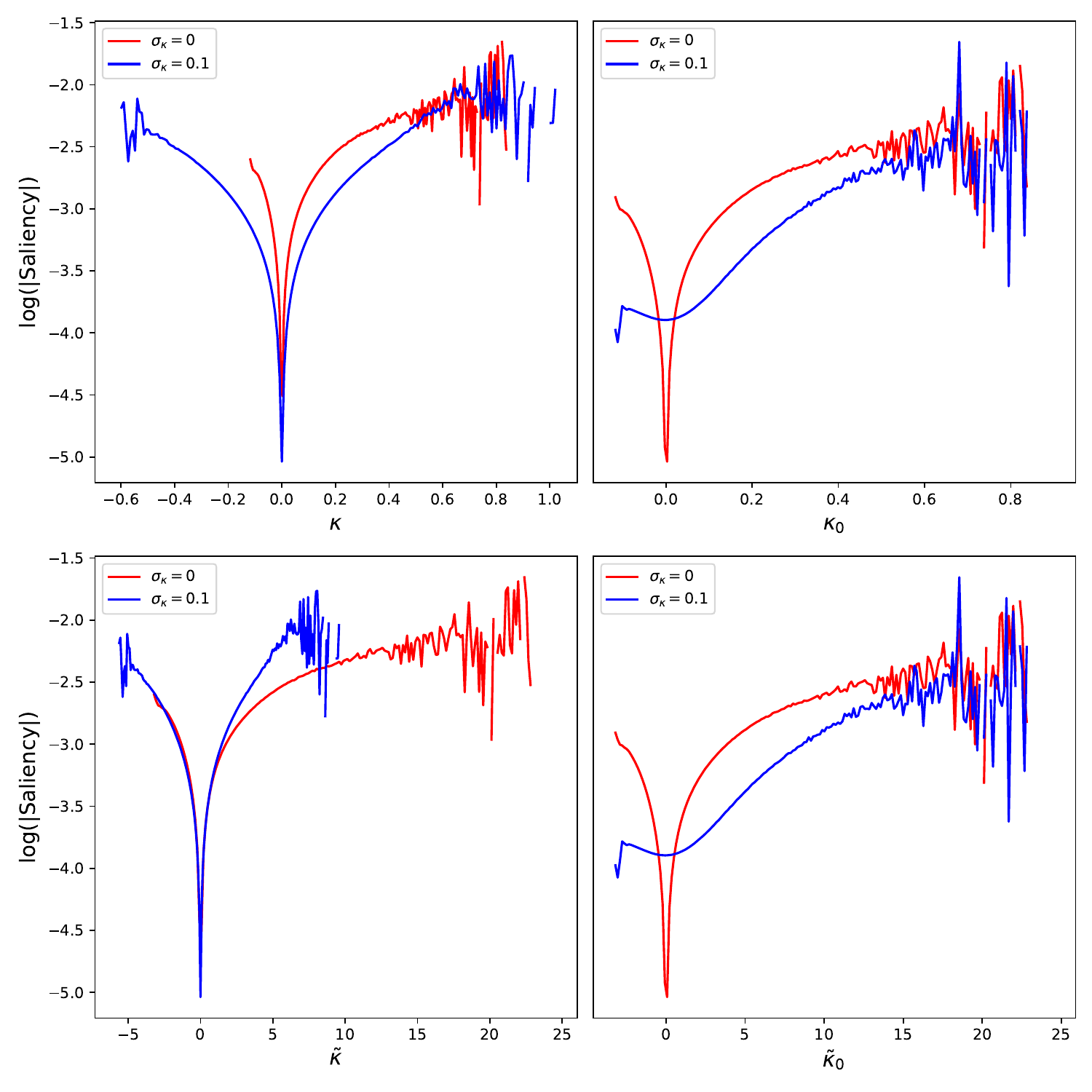}
  \caption{Pixel-averaged values of the logarithm of the absolute value of the Integrated Gradient saliency maps as a function of the convergence map pixel value $\kappa$ (top row) and the standard-deviation-normalized convergence value $\tilde{\kappa}$ (bottom row), for models with $\sigma_\kappa = 0$ (red) and $\sigma_\kappa = 0.1$ (blue). The left column shows results computed using bin edges tailored to each noise level, while the right column compares both noise levels using a shared binning scheme derived from the $\sigma_\kappa = 0$ case. All results correspond to models with $z = 2.5$ and $k_{\text{max}} = \pi/2$, and include both massive and non-massive samples. The 2-point information has not been scaled out.
  Higher saliency values imply that the network is paying more attention to that region of the map, thus the network is paying most attention to extreme values of $\kappa$, i.e.~the most overdense and underdense regions.
  }
  \label{fig:sal_vs_kap}
\end{figure*}

The upper row of Figure \ref{fig:valid_acc} shows the classification accuracy evaluated on the test set for the different scale cuts $k_{\rm max}$, source redshifts $z$, and noise levels $\sigma_\kappa$, as well as the effect of scaling out the 2-point information. It can be seen that the accuracy of the network improves as smaller, nonlinear, scales are included; this is to be expected as the effect of massive neutrinos becomes more prominent on small scales. It can also be seen that the accuracy is higher at higher redshifts; this is to be expected as, the further away the light source is, the further the light travels before reaching the observer, experiencing more lensing and thus producing a larger signal. 
It can also be seen that increasing noise reduces the accuracy, which is to be expected as the massive neutrino signal is distorted by the noise.
The network is able to reach almost $100\%$ accuracy for a source redshift of 2.5 with $\ell_{\rm max}\simeq5{,}000$. However, most interestingly, the accuracy of the network drops to almost $50\%$ after scaling out the 2-point information from the maps for all source redshifts and scales, except for low noise cases at small scales where it reaches $\sim60\%$. This implies that the network is almost unable to tell the difference between the 0eV and 0.1eV maps, suggesting that most of the field-level information can be found using the power spectrum alone. 
One notable exception is in the case of $z=0.5$ where the loss of accuracy from scaling out the 2-point information is relatively small on small scales, implying that -- even though the total information is relatively small (implied by the low accuracy) -- a large fraction of this information exists beyond the 2-point; this is especially the case in the zero-noise limit, implying that futuristic surveys with shape noise levels below LSST will have more information beyond the 2-point.

Following the insight provided by the classification accuracy, we now consider a more statistically interpretable quantity, the likelihood ratio.
Given the use of a binary cross entropy loss function, we can relate the output of the neural network $f(\kappa_{\rm obs})$
to the likelihood ratio between the $M_\nu=0.1{\rm eV}$ and $M_\nu=0$ hypotheses as follows,
\begin{equation}
     e^{\Delta \ln L} \equiv \frac{L(M_\nu=0.1{\rm eV}|\kappa_{\rm obs})}
{L(M_\nu=0|\kappa_{\rm obs})} = \frac{f(\kappa_{\rm obs})}{1-f(\kappa_{\rm obs})},
\end{equation}
where $L(M_\nu|\kappa_{\rm obs})$ is the likelihood and $\kappa_{\rm obs}$ is an observed map from a cosmology with true neutrino mass corresponding to the `1 label' in the classification; here $0.1\,{\rm eV}$ 
\cite{Jeffrey:2023evidenceN}.  
The lower row of Figure \ref{fig:valid_acc} shows how the difference in log likelihood $\Delta \ln L$ varies with $k_{\rm max}$, $z$, and $\sigma_\kappa$, for observations with $M_\nu=0.1\,{\rm eV}$ (taken from the test set).
The overall trends are similar to the classification accuracy, with higher $k_{\rm max}$, higher $z$, and lower $\sigma_\kappa$, leading to a higher likelihood difference. In the most optimistic case of $z=2.5$ and $k_{\rm max}=2\pi\,h/{\rm Mpc}$ $(\ell_{\rm max}\simeq10^4)$ with low noise, the difference in log likelihood is $\Delta \ln L \simeq 4.5$. For a Gaussian likelihood this would correspond to $\Delta\chi^2\simeq9$ and thus a $3\sigma$ detection.\footnote{
We note that, while it is instructive to consider the likelihood ratio to build intuition, this neglects any prior uncertainty on the neutrino mass as it only considers a $M_\nu=0.1{\rm eV}$ alternative hypothesis relative to a $M_\nu=0{\rm eV}$ null hypothesis. In a Bayesian analysis one would marginalize over a prior which spans a range of values of $M_\nu$, carefully taking into account volume effects 
\cite[see e.g.][]{LEE1, Hergt:2021qlh}. This in turn would reduce the significance of any detection of neutrino mass compared to the likelihood analysis here which corresponds to a delta-function prior at the observed value.} This greatly decreases as lower source redshifts, lower scales, or higher noise is considered, dropping to $\Delta \ln L \simeq 0$. Most interestingly, it can also be seen that the likelihood difference is negligible when the 2-point information is removed in \textit{all} cases, implying that there is little information beyond the 2-point.




Figure \ref{fig:saliency} shows saliency maps using the Integrated Gradient method. This enables interpretation of what the network is learning and which parts of the map the model is most sensitive to. 
Each row displays a different weak lensing convergence map in the left column, followed by three saliency maps generated by the best three models from the \texttt{optuna} trials. Higher numbers in the saliency map implies that the network is paying more attention to that region of the map.
We consider the case of $z = 2.5$ and $k_\text{max} = {\pi}/{2}$, comparing models trained with $\sigma_\kappa = 0$ and $\sigma_\kappa = 0.1$. The corresponding saliency maps are expected to be informative. Looking at the Saliency plots for both $\sigma_\kappa = 0$ and $\sigma_\kappa = 0.1$ showed that the networks sometimes looked at both halo and void regions. This implies that there is information in various features of the map, and that using an ensemble approach could allow us to extract more information, although this is beyond the scope of this work.

To quantitatively analyze the saliency, Figure~\ref{fig:sal_vs_kap} shows the pixel-averaged logarithmic saliency as a function of the convergence value $\kappa$ (top) and normalized convergence $\tilde{\kappa}$ (bottom), when binned in terms of the $\kappa$ of each respective map (left) or with respect to the unnoised value $\kappa_0$ (right). In all cases, the saliency is largest in the most underdense and overdense regions of $\kappa$ or $\tilde{\kappa}$. For the noise-free case ($\sigma_\kappa = 0$), there is slightly stronger saliency activation in halo regions, while for the noisy case ($\sigma_\kappa = 0.1$), the activity appears more symmetrically distributed across the two tails. 
This shows the complementarity of cluster-like and void-like regions, as found in 3-dimensional LSS examples which found complementarity between halo and void statistics \cite{bayer2021detecting, Kreisch_2021}. When the saliency is binned in terms of the kappa value of the map in question (left column), the CNN pays equal attention to over and under-dense regions. However, when binning in terms of the unnoised convergence values (right column) the saliency of the noised map is severely washed out. This implies that in the presence of noise, the CNN pays the same attention to over and under-dense regions as it does when there is no noise, however, because many of the most underdense pixels are noise dominated, it is paying attention to noise, and thus information is lost. Underdense regions are less salient in the presence of noise as they have a lower $|\kappa_0|$ compared to overdense regions, and are thus washed out by the noise. This is similar to the trend observed for $\Omega_m$ and $\sigma_8$ by \cite{ZorrillaMatilla:2020doz}, where underdense regions were found to significantly lose information upon the addition of noise, however, our results are different to \cite{ZorrillaMatilla:2020doz} as we find over and underdense regions are similarly salient in the absence of noise, while \cite{ZorrillaMatilla:2020doz} found underdense regions to be more salient -- implying a difference in what is a=most salient for $\Omega_m, \sigma_8,$ and $M_\nu$.

\section{Conclusions}
\label{sec:conclusions}

This work quantified the ability of CNNs to extract information regarding neutrino mass from weak lensing convergence maps.
We studied the information content as a function of scale cut, source redshift, and noise level for an LSST-like survey. We also investigated the effect of scaling out the 2-point information from the maps to study the beyond-2-point information.
We found that a ResNet-18-based classifier was able to distinguish between a 0eV and 0.1eV neutrino mass cosmology with a high accuracy that improves as higher source redshifts and smaller scales are considered, reaching almost $100\%$ for a source redshift of 2.5 with $\ell_{\rm max}\simeq5{,}000$. However, the accuracy of the network dropped to almost $50\%$ after scaling out the 2-point information from the maps, implying that most of the field-level information can be found using the power spectrum alone, in agreement with the simple cross-correlation arguments of \cite{Bayer_2022_fake}.

We also went beyond using the neural network as a black box, by using Integrated Gradient saliency maps to understand which parts of the map the network is learning from. We found that the network pays most attention to the most overdense and underdense regions. In the absence of noise, it pays similar attention to both overdense and underdense regions, with a slight preference for overdense regions. This complements findings from studies that explicitly considered cluster-based and void-based summary statistics and found them to be complementary \cite{bayer2021detecting, Kreisch_2021}. In the presence of noise, the network still pays attention to the most overdense and underdense pixels, however, many of these pixels are noise dominated, thus information is lost. This particularly affects underdense regions, which disproportionately lose information compared to overdense regions. This suggests that while both regions are complementary sources of information, underdense regions are more susceptible to information loss in noisy settings.

In this analysis we were restricted to using a classifier due to the availability of simulations, but in the future, when more simulations are produced, it would be interesting further work to do full parameter inference, studying the beyond 2-point information in a similar manner to \cite{Beyond-2pt:2024mqz} which considered $\Omega_m$ and $\sigma_8$ for halo clustering.
While we have performed extensive hyperparameter tuning to facilitate the optimality of the CNN, it would also be interesting future work to study the field-level neutrino mass information using different architectures and methodologies, such as differentiable forward modelling to reconstruct the initial conditions of the Universe \cite{Jasche_2013, Seljak_2017, Jasche_2019, Schmidt:2020LEFTField, Bohm:2020madlens, Modi_2021, Porqueres:2020wwf, Porqueres:2021clw, Porqueres:2023drp, Boruah:2022lsu, Boruah:2023fph, Boruah:2024tqp, Lanzieri:2024mvn, Zeghal:2024kic, li2022pmwd, Bayer_2023_vel, Bayer:2023mclmc, Zang:2023rpx} and normalizing flows \cite{Hassan:2021ymv, Dai:2022dso, Dai:2023lcb}; although we do not expect this to change the conclusions.

Furthermore, while we considered the information content in weak lensing convergence maps alone, it would be interesting to study the information when cross correlating convergence maps with other probes from LSS and the CMB using cross-survey simulations such as \cite{Omori:2022uox, Bayer:2024hd}, as this could provide a promising source of beyond-2-point information. It would also be fruitful to study the neutrino mass constraints while including systematic effects to investigate whether the relative constraining power of field-level to 2-point is modified. Finally, it would be worthwhile to repeat this analysis while including baryonic effects to study the degeneracy between massive neutrinos and baryons at the field level \cite{Ferlito:2023gum}.


\acknowledgments

We thank Vanessa Böhm, Françoise Lanusse, Jia Liu, Uroš Seljak, Francisco Villaescusa-Navarro, and Benjamin Wandelt for insightful discussion. MG also thanks Vanessa Böhm for her initial mentorship. This research was conducted as part of the N3AS Undergraduate Research Program at UC Berkeley, supported by the NSF Award Number 2020275 and Heising Simons Grant 2017-228. This research used resources of the National Energy Research Scientific Computing Center (NERSC), a Department of Energy Office of Science User Facility using NERSC award ASCR-ERCAP0029232. The computations reported in this paper were (in part) performed using resources made available by the Flatiron Institute. The analysis of the simulations has made use of the \textit{Pylians} library, publicly available at \url{https://github.com/franciscovillaescusa/Pylians3}.


\bibliographystyle{JHEP}
\bibliography{biblio.bib}


\end{document}